\documentstyle[12pt]{article}
\begin{document}
\title{Bound - states for
 spiked harmonic oscillators and truncated Coulomb potentials}
\author{ Omar Mustafa and Maen Odeh\\
 Department of Physics, Eastern Mediterranean University\\
 G. Magusa, North Cyprus, Mersin 10 - Turkey\\
 email: omustafa.as@mozart.emu.edu.tr\\
\date{}\\}
\maketitle
\begin{abstract}
{\small We propose a new analytical method to solve for the nonexactly
solvable Schr\"odinger equation. Successfully, it is applied to a class of
spiked harmonic oscillators and truncated Coulomb potentials. The utility
of this method could be extended to study other systems of atomic,
molecular and nuclear physics interest.}
\end{abstract}
\newpage

In atomic, molecular and nuclear physics, spiked harmonic oscillators
and truncated Coulomb potentials are of significant interest. Realistic
interaction potentials often have a usually repulsive core [1-5]. The
simplest model of such a core is provided by the spiked harmonic
oscillators\\
\begin{equation}
V(q)=c_1 q^2+ c_2 q^{-b},~~~~~~c_1, c_2, b>0.
\end{equation} 
On the other hand, the truncated Coulomb potential has been founded to
be pertinent in the study of the energy levels of the hydrogen - like
atoms exposed to intense laser radiation [6-11]. It has been shown [8,10]
that under Kramers - Henneberger transformation [12] the laser - dressed
binding potential for the hydrogenic system, often called laser - dressed
Coulomb potential, may be well simulated by\\
\begin{equation}
V(q)=-\frac{e^2}{(q^2+c^2)^{1/2}}~~;~~~~~~c>0.
\end{equation} 
Where the truncation parameter $c$ is related to the strength of the
irradiating laser field.

Thus it is interesting to carry out systematic studies of the bound - states
of these potentials. Hall and Saad [4] have studied the spiked harmonic
oscillator potentials via smooth transformations method (STM) of the exactly
solvable potential $V(q)=c_2q^2+c_2q^{-2}$ to obtain lower and/or upper
energy bounds. They have also calculated the energy eigenvalues using
direct numerical integrations of Schr\"odinger equation [4]. Dutt et al [6]
have used a shifted 1/N expansion technique (SLNT) to carry out the
energy levels of the laser - dressed Coulomb potential and compared their
results also with those of direct numerical integrations [7]. Nevertheless,
neither SLNT nor STM is utilitarian in terms of calculating the eigenvalues
and eigenfunctions in one batch. Because of the complexity in handling
large - order corrections of the standard Rayleigh - Schr\"odinger
perturbation theory, only low - order calculations have been reported
for SLNT [6,13] and large - order calculations have been neglected.
Eventually, the results of SLNT are not as accurate as sought after.

In this paper we formulate a method for solving Schr\"odinger equation. In
one batch, one should be able to study not only the eigenvalues but also the
eigenfunctions. It simply consists of using $1/\bar{l}$ as a perturbation
expansion parameter. Where, $\bar{l}=l-\beta$, $l$ is a quantum number, and
$\beta$ is a suitable shift introduced to avoid the trivial case $l=0$.
Hence, hereinafter, it should be called pseudoperturbative shifted - $l$
expansion technique (PSLET).

The construction of our method starts with the time - independent 
one - dimensional form of Schr\"odinger equation, in 
$\hbar = m = 1$ units,\\
\begin{equation}
\left[-\frac{1}{2}\frac{d^{2}}{dq^{2}}+\frac{l(l+1)}{2q^{2}}+V(q)\right]
\Psi_{n_r,l}(q)=E_{n_r,l}\Psi_{n_r,l}(q).
\end{equation} 
Where the quantum number $l$ may specify parity, $(-1)^{l+1}$, in 
one - dimension ( $l$=-1 or $l$=0, and $q \in ( -\infty, \infty)$) or
angular momentum in three - dimensions ( $l$=0, 1, $\cdots$ , and 
$q \in (0, \infty)$), and $n_r$=0, 1, $\cdots$ counts the nodal zeros 
[6,14-17].

To avoid the trivial case $l$=0, the quantum number $l$ is shifted
through the relation $\bar{l} = l - \beta$. Eq.(3) thus becomes\\
\begin{equation}
\left\{-\frac{1}{2}\frac{d^{2}}{dq^{2}}+\tilde{V}(q)\right\}
\Psi_{n_r,l} (q)=E_{n_r,l}\Psi_{n_r,l}(q),
\end{equation} 
\begin{equation}
\tilde{V}(q)=\frac{\bar{l}^{2}+(2\beta+1)\bar{l}
+\beta(\beta+1)}{2q^{2}}+\frac{\bar{l}^2}{Q}V(q).
\end{equation} 
Herein, it should be noted that Q is a constant that scales the potential 
$V(q)$ at large - $l$ limit and is set, for any specific choice of $l$
and $n_r$, equal to $\bar{l}^2$ at the end of the calculations [13,14].
And, $\beta$ is to be determined in the sequel.

Our systematic procedure begins with shifting the origin of the
coordinate through\\
\begin{equation}
x=\bar{l}^{1/2}(q-q_{o})/q_{o},
\end{equation} 
where $q_{o}$ is currently an arbitrary point to perform Taylor expansions
about, with its particular value to be determined. Expansions about
this point, $x=0$ (i.e. $q=q_o$), yield\\
\begin{equation}
\frac{1}{q^{2}}=\sum^{\infty}_{n=0} (-1)^{n} \frac{(n+1)}{q_{o}^{2}}
 x^{n}\bar{l}^{-n/2},
\end{equation} 
\begin{equation}
V(x(q))=\sum^{\infty}_{n=0}\left(\frac{d^{n}V(q_{o})}{dq_{o}^{n}}\right)
\frac{(q_{o}x)^{n}}{n!}\bar{l}^{-n/2}.
\end{equation} 
Obviously, the expansions in (7) and (8) center the problem at an
arbitrary point $q_o$ and the derivatives, in effect, contain
information not only at $q_o$ but also at any point on the axis, in
accordance with Taylor's theorem. Also it should be mentioned here
that the scaled coordinate, equation (6), has no effect on the energy
eigenvalues, which are coordinate - independent. It just facilitates
the calculations of both the energy eigenvalues and eigenfunctions.
It is also convenient to expand $E$ as\\
\begin{equation}
E_{n_r,l}=\sum^{\infty}_{n=-2}E_{n_r,l}^{(n)}\bar{l}^{-n}.
\end{equation} 
Equation (4) thus becomes\\
\begin{equation}
\left[-\frac{1}{2}\frac{d^{2}}{dx^{2}}+\frac{q_{o}^{2}}{\bar{l}}
\tilde{V}(x(q))\right]
\Psi_{n_r,l}(x)=\frac{q_{o}^2}{\bar{l}}E_{n_r,l}\Psi_{n_r,l}(x),
\end{equation} 
with\\
\begin{eqnarray}
\frac{q_o^2}{\bar{l}}\tilde{V}(x(q))&=&q_o^2\bar{l}
\left[\frac{1}{2q_o^2}+\frac{V(q_o)}{Q}\right]
+\bar{l}^{1/2}\left[-x+\frac{V^{'}(q_o)q_o^3 x}{Q}\right]\nonumber\\
&+&\left[\frac{3}{2}x^2+\frac{V^{''}(q_o) q_o^4 x^2}{2Q}\right]
+(2\beta+1)\sum^{\infty}_{n=1}(-1)^n \frac{(n+1)}{2}x^n \bar{l}^{-n/2}
\nonumber\\
&+&q_o^2\sum^{\infty}_{n=3}\left[(-1)^n \frac{(n+1)}{2q_o^2}x^n
+\left(\frac{d^n V(q_o)}{dq_o^n}\right)\frac{(q_o x)^n}{n! Q}\right]
\bar{l}^{-(n-2)/2}\nonumber\\
&+&\beta(\beta+1)\sum^{\infty}_{n=0}(-1)^n\frac{(n+1)}{2}x^n
\bar{l}^{-(n+2)/2}+\frac{(2\beta+1)}{2},
\end{eqnarray}\\
where the prime of $V(q_o)$ denotes derivative with respect to $q_o$.
Equation (10) is exactly of the type of Schr\"odinger equation 
for one - dimensional anharmonic oscillator\\
\begin{equation}
\left[-\frac{1}{2}\frac{d^2}{dx^2}+\frac{1}{2}w^2 x^2 +\varepsilon_o
+P(x)\right]X_{n_r}(x)=\lambda_{n_r}X_{n_r}(x),
\end{equation} 
where $P(x)$ is a perturbation - like term and $\varepsilon_o$ is a 
constant. A simple comparison between Eqs.(10), (11) and (12) implies\\
\begin{equation}
\varepsilon_o =\bar{l}\left[\frac{1}{2}+\frac{q_o^2 V(q_o)}{Q}\right]
+\frac{2\beta+1}{2}+\frac{\beta(\beta+1)}{2\bar{l}},
\end{equation} 
\begin{eqnarray}
\lambda_{n_{r}}&=&\bar{l}\left[\frac{1}{2}+\frac{q_o^2 V(q_o)}{Q}\right]
+\left[\frac{2\beta+1}{2}+(n_r+\frac{1}{2})w\right]\nonumber\\
&+&\frac{1}{\bar{l}}\left[\frac{\beta(\beta+1)}{2}+\lambda_{n_{r}}^{(0)}\right]
+\sum^{\infty}_{n=2}\lambda_{n_{r}}^{(n-1)}\bar{l}^{-n},
\end{eqnarray}\\
and\\
\begin{equation}
\lambda_{n_{r}} = q_o^2 \sum^{\infty}_{n=-2} E_{n_r,l}^{(n)}
\bar{l}^{-(n+1)},
\end{equation} 
Equations (14) and (15) yield\\
\begin{equation}
E_{n_r,l}^{(-2)}=\frac{1}{2q_o^2}+\frac{V(q_o)}{Q}
\end{equation} 
\begin{equation}
E_{n_r,l}^{(-1)}=\frac{1}{q_o^2}\left[\frac{2\beta+1}{2}
+(n_r +\frac{1}{2})w\right]
\end{equation} 
\begin{equation}
E_{n_r,l}^{(0)}=\frac{1}{q_o^2}\left[ \frac{\beta(\beta+1)}{2}
+\lambda_{n_r}^{(0)}\right]
\end{equation} 
\begin{equation}
E_{n_r,l}^{(n)}=\lambda_{n_r}^{(n)}/q_o^2  ~~;~~~~n \geq 1.
\end{equation} 
Here $q_o$ is chosen to minimize $E_{n_r,l}^{(-2)}$, i. e.\\
\begin{equation}
\frac{dE_{n_r,l}^{(-2)}}{dq_o}=0~~~~
and~~~~\frac{d^2 E_{n_r,l}^{(-2)}}{dq_o^2}>0.
\end{equation} 
Hereby, $V(q)$ is assumed to be well behaved so that $E^{(-2)}$ has
a minimum $q_o$ and there are well - defined bound - states.
Equation (20) in turn gives, with $\bar{l}=\sqrt{Q}$,\\
\begin{equation}
l-\beta=\sqrt{q_{o}^{3}V^{'}(q_{o})}.
\end{equation} 
Consequently, the second term in Eq.(11) vanishes and the first term adds 
a constant to the energy eigenvalues. It should be noted that energy term $\bar{l}^2E_{n_r,l}^{(-2)}$ has its counterpart in classical
mechanics. It corresponds roughly to the energy of a classical particle
with angular momentum $L_z$=$\bar{l}$  executing circular motion  of 
radius $q_o$ in the potential $V(q_o)$. This term thus identifies the 
leading - order approximation, to all eigenvalues, as a classical 
approximation and the higher - order corrections as quantum fluctuations
around the minimum $q_o$, organized in inverse powers of $\bar{l}$.

The next leading correction to the energy series, $\bar{l}E_{n_r,l}^{(-1)}$,
consists of a constant term and the exact eigenvalues of the unperturbed
harmonic oscillator potential $w^2x^2/2$. 
The shifting parameter $\beta$ is determined by choosing
$\bar{l}E_{n_r,l}^{(-1)}$=0. This choice is physically motivated. It requires
not only the agreements between PSLET eigenvalues and the exact known ones for
the harmonic oscillator and Coulomb potentials but also between the
eigenfunctions as well.  Hence\\
\begin{equation}
\beta=-\left[\frac{1}{2}+(n_{r}+\frac{1}{2})w\right],
\end{equation} 
where\\
\begin{equation}
w=\sqrt{3+\frac{q_o V^{''}(q_o)}{V^{'}(q_o)}}.
\end{equation} 

Then equation (11) reduces to\\
\begin{equation}
\frac{q_o^2}{\bar{l}}\tilde{V}(x(q))=
q_o^2\bar{l}\left[\frac{1}{2q_o^2}+\frac{V(q_o)}{Q}\right]+
\sum^{\infty}_{n=0} v^{(n)}(x) \bar{l}^{-n/2},
\end{equation} 
where\\
\begin{equation}
v^{(0)}(x)=\frac{1}{2}w^2 x^2 + \frac{2\beta+1}{2},
\end{equation} 
\begin{equation}
v^{(1)}(x)=-(2\beta+1) x - 2x^3 + \frac{q_o^5 V^{'''}(q_o)}{6 Q} x^3,
\end{equation} 
and for $n \geq 2$\\
\begin{eqnarray}
v^{(n)}(x)&=&(-1)^n (2\beta+1) \frac{(n+1)}{2} x^n
+ (-1)^{n} \frac{\beta(\beta+1)}{2} (n-1) x^{(n-2)}\nonumber\\
&+& \left[(-1)^{n} \frac{(n+3)}{2}
+ \frac{q_o^{(n+4)}}{Q(n+2)!} \frac{d^{n+2} V(q_o)}{dq_o^{n+2}}\right]
x^{n+2}.
\end{eqnarray}\\
Equation (10) thus becomes\\
\begin{eqnarray}
&&\left[-\frac{1}{2}\frac{d^2}{dx^2} + \sum^{\infty}_{n=0} v^{(n)}
\bar{l}^{-n/2}\right]\Psi_{n_r,l} (x)= \nonumber\\
&& \left[\frac{1}{\bar{l}}\left(\frac{\beta(\beta+1)}{2}
+\lambda_{n_r}^{(0)}\right) 
+ \sum^{\infty}_{n=2} \lambda_{n_r}^{(n-1)}
\bar{l}^{-n} \right] \Psi_{n_r,l}(x).
\end{eqnarray}\\

When setting the nodeless, $n_r = 0$, wave functions as \\
\begin{equation}
\Psi_{0,l}(x(q)) = exp(U_{0,l}(x)),
\end{equation} 
equation (28) is readily transformed into the following Riccati equation:\\
\begin{eqnarray}
-\frac{1}{2}[ U^{''}(x)+U^{'}(x)U^{'}(x)]
+\sum^{\infty}_{n=0} v^{(n)}(x) \bar{l}^{-n/2}
&=&\frac{1}{\bar{l}} \left( \frac{\beta(\beta+1)}{2}
+ \lambda_{0}^{(0)}\right)\nonumber\\
&&+\sum^{\infty}_{n=2} \lambda_{0}^{(n-1)} \bar{l}^{-n}.
\end{eqnarray}\\
Hereinafter, we shall use $U(x)$ instead of $U_{0,l}(x)$ for simplicity,
and the prime of $U(x)$ denotes derivative with respect to $x$. It is
evident that this equation admits solution of the form \\
\begin{equation}
U^{'}(x)=\sum^{\infty}_{n=0} U^{(n)}(x) \bar{l}^{-n/2}
+\sum^{\infty}_{n=0} G^{(n)}(x) \bar{l}^{-(n+1)/2},
\end{equation} 
where\\
\begin{equation}
U^{(n)}(x)=\sum^{n+1}_{m=0} D_{m,n} x^{2m-1} ~~~~;~~~D_{0,n}=0,
\end{equation} 
\begin{equation}
G^{(n)}(x)=\sum^{n+1}_{m=0} C_{m,n} x^{2m}.
\end{equation} 
Substituting equations (31) - (33) into equation (30) implies\\
\begin{eqnarray}
&-&\frac{1}{2} \sum^{\infty}_{n=0}\left[U^{(n)^{'}} \bar{l}^{-n/2}
+ G^{(n)^{'}} \bar{l}^{-(n+1)/2}\right] \nonumber\\
&-&\frac{1}{2} \sum^{\infty}_{n=0} \sum^{\infty}_{p=0}
\left[ U^{(n)}U^{(p)} \bar{l}^{-(n+p)/2}
+G^{(n)}G^{(p)} \bar{l}^{-(n+p+2)/2}
+2 U^{(n)}G^{(p)} \bar{l}^{-(n+p+1)/2}\right]\nonumber\\
&+&\sum^{\infty}_{n=0}v^{(n)} \bar{l}^{-n/2}
=\frac{1}{\bar{l}}\left(\frac{\beta(\beta+1)}{2}+\lambda_{0}^{(0)}\right)
+\sum^{\infty}_{n=2} \lambda_{0}^{(n-1)} \bar{l}^{-n},
\end{eqnarray}\\
where primes of $U^{(n)}(x)$ and $G^{(n)}(x)$ denote derivatives
with respect to $x$. Equating the coefficients of the same powers of
$\bar{l}$ and $x$, respectively, ( of course the other way around would 
work equally well) one obtains\\
\begin{equation}
-\frac{1}{2}U^{(0)^{'}} - \frac{1}{2}  U^{(0)} U^{(0)} + v^{(0)} = 0,
\end{equation} 
\begin{equation}
U^{(0)^{'}}(x) = D_{1,0} ~~~;~~~~D_{1,0}=-w,
\end{equation} 
and integration over $dx$ yields\\
\begin{equation}
U^{(0)}(x)=-wx.
\end{equation} 
Similarly,\\
\begin{equation}
-\frac{1}{2}[U^{(1)^{'}} + G^{(0)^{'}}] - U^{(0)}U^{(1)} - U^{(0)}G^{(0)}
+v^{(1)}=0,
\end{equation} 
\begin{equation}
U^{(1)}(x)=0,
\end{equation} 
\begin{equation}
G^{(0)}(x)=C_{0,0}+C_{1,0}x^2,
\end{equation} 
\begin{equation}
C_{1,0}=-\frac{B_{1}}{w},
\end{equation} 
\begin{equation}
C_{0,0}=\frac{1}{w}(C_{1,0}+2\beta+1),
\end{equation} 
\begin{equation}
B_{1}=-2+\frac{q_o^5}{6Q}\frac{d^3 V(q_o)}{dq_o^3},
\end{equation} 
\begin{eqnarray}
&&-\frac{1}{2}[U^{(2)^{'}} + G^{(1)^{'}}]
- \frac{1}{2}\sum^{2}_{n=0}U^{(n)}U^{(2-n)}-\frac{1}{2}G^{(0)}G^{(0)}
\nonumber\\
&&-\sum^{1}_{n=0}U^{(n)}G^{(1-n)} + v^{(2)}
= \frac{\beta(\beta+1)}{2} + \lambda_{0}^{(0)},
\end{eqnarray}\\
\begin{equation}
U^{(2)}(x)=D_{1,2}x + D_{2,2}x^3,
\end{equation} 
\begin{equation}
G^{(1)}(x)=0,
\end{equation} 
\begin{equation}
D_{2,2}=\frac{1}{w}(\frac{C_{1,0}^2}{2}-B_{2})
\end{equation} 
\begin{equation}
D_{1,2}=\frac{1}{w}(\frac{3}{2}D_{2,2}+C_{0,0}C_{1,0}
-\frac{3}{2}(2\beta+1)),
\end{equation} 
\begin{equation}
B_{2}=\frac{5}{2}+\frac{q_o^6}{24Q}\frac{d^4V(q_o)}{dq_o^4},
\end{equation} 
\begin{equation}
\lambda_{0}^{(0)} = -\frac{1}{2}(D_{1,2}+C_{0,0}^2).
\end{equation} 
$\cdots$ and so on. Thus, one can calculate the energy 
eigenvalue and the eigenfunctions from the knowledge of $C_{m,n}$
and $D_{m,n}$ in a hierarchical manner.
Nevertheless, the procedure just described is suitable for systematic calculations
using software packages (such as MATHEMATICA, MAPLE, or REDUCE) to determine
the energy eigenvalue and eigenfunction corrections up to any order of the
pseudoperturbation series. 

Although the energy series, Eq.(9), could appear
divergent, or, at best, asymptotic for small $\bar{l}$, one can still 
calculate the eigenenergies to a very good accuracy by forming the 
sophisticated Pade' approximation to the energy series. 
The energy series, Eq.(9), is calculated up to 
$E_{0,l}^{(4)}/\bar{l}^4$ by
\begin{equation}
E_{0,l}=\bar{l}^{2}E_{0,l}^{(-2)}+E_{0,l}^{(0)}+\cdots
+E_{0,l}^{(4)}/\bar{l}^4+O(1/\bar{l}^5),
\end{equation} 
and with the $P_{3}^{3}(1/\bar{l})$ and $P_{3}^{4}(1/\bar{l})$
Pade' approximants it becomes\\
\begin{equation}
E_{0,l}[3,3]=\bar{l}^{2}E_{0,l}^{(-2)}+P_{3}^{3}(1/\bar{l}).
\end{equation} 
and\\
\begin{equation}
E_{0,l}[3,4]=\bar{l}^{2}E_{0,l}^{(-2)}+P_{3}^{4}(1/\bar{l}).
\end{equation} 
Hereby, an " if " statement is in point. If the energy series, eq.(9), is a
Stieltjes series, though it is difficult to prove, then $E_{0,l}[3,3]$
and $E_{0,l}[3,4]$ provide upper and lower bounds to the energy [18,19].
Our strategy is therefore clear.

Let us begin with the spiked harmonic oscillators\\
\begin{equation}
V(q)=\frac{1}{2}(q^2+aq^{-b})
\end{equation} 
for which Eq.(22), with $n_r=0$, implies\\
\begin{equation}
\beta=-\frac{1}{2}(1+w)~~;~~~~w=\sqrt{\frac{8q_o+ab(b-2)q_o^{-(b+1)}}
{2q_o-abq_o^{-(b+1)}}}.
\end{equation} 
In turn Eq.(21) reads\\
\begin{equation}
l+\frac{1}{2}\left(1+\sqrt{\frac{8q_o+ab(b-2)q_o^{-(b+1)}}
{2q_o-abq_o^{-(b+1)}}}\right)
=q_o^2 \sqrt{1-\frac{ab}{2}q_o^{-(b+2)}}.
\end{equation} 
Equation (56) is explicit in $q_o$ and evidently a closed form solution
for $q_o$ is hard to find, though almost impossible. However, numerical
solutions are feasible. Once $q_o$ is determined the coefficients
$C_{m,n}$ and $D_{m,n}$ are obtained in a sequel manner. Consequently, the
eigenvalues, Eq.(51), and eigenfunctions, Eqs.(31)-(33), are calculated
in the same batch for each value of $a$, $b$, and $l$. In tables 1 and 2
we list PSLET results $E_P$, Eq.(51), along with [3,3] and [3,4] Pad\'{e}
approximants, Eqs.(52) and (53) respectively. The results of the smooth
transformations method (STM) [4] and direct numerical integration (DNI) [4]
are also displayed for comparison purposes.

Our calculated values of the bound - state energies, $E_P$, compare well
with those from direct numerical integrations [4]. In table 1 the Pad\'{e}
approximants $E[3,3]$ and $E[3,4]$ are almost in total agreement with
those of Hall and Saad [4] from DNI of the Schr\"odinger equation. Moreover,
it is evident that $E[3,3]$ and $E[3,4]$ have provided upper and lower
bounds, respectively, to the energy series. However, the same can not be
concluded from table 2. Eventually, our computed values of the bound - state
energies, $E_P$, do not contradict with the upper and/or lower bounds
reported by Hall and Saad [4] from the smooth transformations method (STM).

Moreover, our result for $b=2$ listed in table 1 is in excellent agreement
with the exact one 65.2534584 obtained from Eq.(2) of ref.[4]. On the other
hand, one would rewrite the centrifugal term in (3) plus the potential (54)
as $l'(l'+1)/(2q^2)+q^2/2$, where $l'=-1/2+\sqrt{(l+1/2)^2+a}$, and proceed
by shifting the irrational quantum number $l'$ through
$\bar{l}=l'-\beta$. In this case, one obtains the known exact result
$E_P=(l'+3/2)$ for the harmonic oscillator $q^2/2$ from the leading term
$\bar{l}^2E^{(-2)}$ and the remainder energy corrections are identically
zero.

Next, we consider the laser - dressed Coulomb potential\\
\begin{equation}
V(q)=-\frac{1}{\sqrt{q^2+c^2}}~~,~~~~c>0.
\end{equation} 
In this case\\
\begin{equation}
w=\sqrt{\frac{q_o^2+4c^2}{q_o^2+c^2}},
\end{equation} 
and\\
\begin{equation}
l+\frac{1}{2}\left(1+\sqrt{\frac{q_o^2+4c^2}{q_o^2+c^2}}\right)
=q_o^2\left[q_o^2+c^2\right]^{-3/4}.
\end{equation} 
Again, we  numerically solve for $q_o$ and proceed exactly as above to
calculate the energy eigenvalues and eigenfunctions in the same batch.
In tables 3 and 4 we collect the results for the truncation parameter
$c=1,5,10,50,100,200$ based on our approach. The energies $E_P$, Eq.(51),
compare well with those of Singh et al. [7] from numerical integrations.
The Pad\'{e} approximants $E[3,3]$ and $E[3,4]$ are in almost  complete
accord with those of Singh et al.[7]. However, they do not provide upper
and lower bounds to the energy series, Eq.(51). Perhaps, it should be
mentioned that the approximate binding potential Eq.(57) is valid for a
hydrogen atom in a laser field which corresponds to a truncation parameter
$c$ in the range 20-60 [6]. Higher and lower values of $c$ have been
considered for academic interest only.

Before we conclude some remarks deserve to be mentioned.

For the two problems discussed in this paper, we  have shown that it is an
easy task to implement PSLET without having to worry about the ranges of
couplings and forms of perturbations in the potential involved.
In contrast to the textbook Rayleigh - Schr\"odinger
perturbation theory, an easy feasibility of
computation of the eigenvalues and eigenfunctions, in one batch, 
has been demonstrated, and satisfactory accuracies have been obtained. 
Moreover, a nice numerical trend of convergence has been achieved.
Nevertheless, another suitable criterion for choosing the value
of the shift $\beta$, reported in Ref. [14], is also feasible. This
reference should be consulted for more details.

It is not easy to prove that the energy series Eq.(51) is a Stieltjes
series. But, if it is a Stieltjes series, the $[N,N]$ and $[N,N+1]$
Pad\'{e} approximants provide upper and lower bounds to the energy series.
Table 1 bears this out. Moreover, in view of the results listed in
tables 1-4 one can confidently conclude that the [3,3] and [3,4] Pad\'{e}
approximants to the energy series Eq.(51) can be used to determine the
energy eigenvalues to a very satisfactory accuracy.

From the knowledge of $C_{m,n}$ and $D_{m,n}$ one can calculate, in
the same batch, the wave functions to study electronic transitions and
multiphoton emission occurring in atomic systems in the presence of
intense laser fields, for example. Such studies already lie beyond the
scope of our present methodical proposal.

Finally, the attendant technique PSLET could be applied to Schr\"odinger
equation with rational potentials, such as the nonpolynomial oscillator
$V(q)=q^2+\lambda q^2/(1+gq^2)$. This type of potential is an interesting
model in laser and quantum field theories [20]. The feasibility of PSLET
extends also to a class of screened Coulomb potentials, which have relevance
in atomic and  plasma physics, and to some other models of interest
[21-26, and references therein].

\newpage
\begin{table}
\begin{center}
\caption{ 1s - state energies, in $\hbar=m=1$ units, of the potential
$V(q)=(q^2+1000/q^b)/2$. Where $E_P$ represents PSLET results, Eq.(51),
$E_S^{U,L}$ with $U$ and $L$ denote upper and lower bounds from STM [4] 
, and $E_N$ from DNI [4].
$E[3,4]$ is the [3,4] Pad\'{e} approximant obtained by replacing
the last $j$ digits of $E[3,3]$ with the $j$ digits in parentheses.}
\vspace{1cm}
\begin{tabular}{|ccccc|}
\hline\hline
$b$ & $E_P$ & $E[3,3]~\&~(E[3,4])$ & $E_S$ & $E_N$\\
\hline
0.5 & 415.88978  & 415.889786 (86)  & 416.30977$^U$ & 415.88979\\
1.0 & 190.72330  & 190.723308 (07)  & 190.99213$^U$ & 190.72331\\
1.5 & 104.41022  & 104.410224 (24) & 104.53993$^U$ & 104.41022\\
1.9 & 71.06157   & 71.0615789 (87)  & 71.08686$^U$  & 71.06158\\
2.0 & 65.25345   & 65.2534589 (86)  & 65.25346      & 65.25346\\
2.1 & 60.15200   & 60.1520114 (11)  & 60.12704$^L$  & 60.15201\\
2.5 & 44.95547   & 44.9554855 (50)  & 44.83349$^L$  & 44.95549\\
3.0 & 33.31675   & 33.3167621 (18)  & 33.07940$^L$  & 33.31676\\
3.5 & 26.10884   & 26.1088462 (48)  & 25.76204$^L$  & 26.10885\\
4.0 & 21.36950   & 21.3694640 (14)  & 20.91865$^L$  & 21.36964\\
4.5 & 18.10194   & 18.1018377 (10)  & 17.55218$^L$  & 18.10183\\
5.0 & 15.76134   & 15.761144  (25)  & 15.11758$^L$  & 15.76113\\
5.5 & 14.03138   & 14.03112   (07)  & 13.29842$^L$  & 14.03107\\
6.0 & 12.71886   & 12.71879   (61)  & 11.90153$^L$  & 12.71862\\
\hline\hline
\end{tabular}
\end{center}
\end{table}
\newpage
\begin{table}
\begin{center}
\caption{ 1s - state energies, in $\hbar=m=1$ units, of the potential
$V(q)=(q^2+a/q^{5/2})/2$. Where $E_P$ represents PSLET results, Eq.(51),
$E_S$ denotes the lower bounds from STM [4] 
, and $E_N$ from DNI [4].
$E[3,4]$ is the [3,4] Pad\'{e} approximant obtained by replacing
the last $j$ digits of $E[3,3]$ with the $j$ digits in parentheses.}
\vspace{1cm}
\begin{tabular}{|ccccc|}
\hline\hline
$a$ & $E_P$ & $E[3,3]~\&~(E[3,4])$ & $E_S$ & $E_N$\\
\hline
1000 & 44.95547 & 44.9554855 (50)  & 44.83349 & 44.95549\\
100  & 17.54168 & 17.541911 (899)  & 17.41900 & 17.54189\\
10   & 7.73423  & 7.73606   (548)  & 7.61169  & 7.73511\\
5    & 6.29679  & 6.29988   (756)  & 6.17394  & 6.29647\\
1    & 4.32861  & 4.528     (290)  & 4.20453  & 4.31731\\
0.5  & 3.85740  & 3.8308    (289)  & 3.74611  & 3.84855\\
0.05 & 3.13431  & 3.1606    (893)  & 3.10954  & 3.15243\\
0.005& 3.01445  & 3.0199    (201)  & 3.01178  & 3.01905\\
\hline\hline
\end{tabular}
\end{center}
\end{table}
\newpage
\begin{table}
\begin{center}
\caption{ Bound - state energies, in $\hbar=m=1$ units, of the potential
$V(q)=-(q^2+c^2)^{-1/2}$ for the 1s, 2p, 3d, and 4f states.
Where $E_P$ represents PSLET results Eq.(51),
$E_{SLNT}$ from SLNT [6], and $E_N$ from DNI [7].
$E[3,4]$ is the [3,4] Pad\'{e} approximant obtained by replacing
the last $j$ digits of $E[3,3]$ with the $j$ digits in parentheses.}
\vspace{1cm}
\begin{tabular}{|cccccc|}
\hline\hline
$c$ & State & $-E_P$ & $-E[3,3]~\&~(-E[3,4])$ & $-E_{SLNT}$ & $-E_N$\\
\hline
1 & 1s & 0.27412    & 0.27478 (62)   & 0.27596  & 0.27439\\
  & 2p & 0.113087   & 0.11296 (303)  & 0.112826 & 0.113024\\
  & 3d & 0.0544357  & 0.0544371 (82) & 0.054442 & 0.0544362\\
  & 4f & 0.03106845 & 0.03106846 (47)& 0.031069 & 0.03106846\\
\hline
5 & 1s & 0.1070836  & 0.1070813 (10) & 0.107396 & 0.1070814\\
  & 2p & 0.06819140 & 0.06818667 (33)& 0.068233 & 0.06818716\\
  & 3d & 0.04325586 & 0.04325730 (20)& 0.043247 & 0.04325755\\
  & 4f & 0.02810534 & 0.02810520 (25)& 0.028101 & 0.02810524\\
\hline
10& 1s & 0.06373831 & 0.06373817 (21)& 0.063820 & 0.0637389\\
  & 2p & 0.04620043 & 0.04619903 (00)& 0.046228 & 0.04619904\\
  & 3d & 0.03315868 & 0.03315855 (53)& 0.033164 & 0.03315859\\
  & 4f & 0.02380662 & 0.02380672 (71)& 0.023806 & 0.02380674\\
\hline\hline
\end{tabular}
\end{center}
\end{table}
\newpage
\begin{table}
\begin{center}
\caption{ Bound - state energies, in $\hbar=m=1$ units, of the potential
$V(q)=-(q^2+c^2)^{-1/2}$ for the 1s, 2p, 3d, and 4f states.
Where $E_P$ represents PSLET results Eq.(51),
$E_{SLNT}$ from SLNT [6], and $E_N$ from DNI [7].
$E[3,4]$ is the [3,4] Pad\'{e} approximant obtained by replacing
the last $j$ digits of $E[3,3]$ with the $j$ digits in parentheses.}
\vspace{1cm}
\begin{tabular}{|cccccc|}
\hline\hline
$c$ & State & $-E_P$ & $-E[3,3]~\&~(-E[3,4])$ & $-E_{SLNT}$ & $-E_N$\\
\hline
50 & 1s & 0.01626071 & 0.0.01626072 (71) & 0.016263 & 0.01626072\\
   & 2p & 0.01408837 & 0.01408837 (37)  & 0.014090 & 0.01408838\\
   & 3d & 0.01215871 & 0.01215871 (71) & 0.012160 & 0.01215871\\
   & 4f & 0.01045842 & 0.01045842 (42) & 0.010459 & 0.01045842\\
\hline
100 & 1s & 0.00862978  & 0.00862978 (78) & 0.008630 & 0.00862978\\
    & 2p & 0.00780013 & 0.00780013 (13)& 0.007800 & 0.00780013\\
    & 3d & 0.00703519 & 0.00703519 (19)& 0.007035 & 0.00703519\\
    & 4f & 0.00633273 & 0.00633273 (73)& 0.006333 & 0.00633273\\
\hline
200& 1s & 0.00450285 & 0.00450285 (85)& 0.004503 & 0.00450286\\
   & 2p & 0.00419307 & 0.00419307 (07)& 0.004193 & 0.00419307\\
   & 3d & 0.00390020 & 0.00390020 (20)& 0.003900 & 0.00390020\\
   & 4f & 0.00362385 & 0.00362385 (85)& 0.003624 & 0.00362385\\
\hline\hline
\end{tabular}
\end{center}
\end{table}


\newpage
\begin{thebibliography}{99}
\bibitem {} Znojil M 1992 Phys Lett {\bf A169} 415
\bibitem {} Flynn M F, Guardiola R and Znojil M 1991 Czech J Phys
           {\bf B41} 1019
\bibitem {} Aguilera V C, Estevez G A and Guardiola R 1990 J Math Phys
           {\bf 31} 99
\bibitem {} Hall R L and Saad N 1998 J Phys {\bf A31} 963
\bibitem {} Hall R L, Saad N and Kevicziky R B 1998
            J Math Phys {\bf 39} 6345
\bibitem {} Dutt R, Mukherji U and Varshni Y P 1985
            J Phys {\bf B18} 3311 (and references therein)
\bibitem {} Singh D and Varshni Y P 1985 Phys Rev {\bf A32} 619
\bibitem {} Miranda L C 1981 Phys Lett {\bf A86} 363
\bibitem {} Lima C A and Miranda L C 1981 Phys Lett {\bf A86} 367
\bibitem {} Lima C A and Miranda L C 1981 Phys Rev {\bf A23} 3335
\bibitem {} Landgraf T C et al. 1982 Phys Lett {\bf A92} 131
\bibitem {} Henneberger W C 1968 Phys Rev Lett {\bf 21} 838
\bibitem {} Imbo T, Pagnamenta A and Sukhatme U 1984
            Phys Rev {\bf D29} 1669
\bibitem {} Maluendes S A, Fernandez F M and Castro E A 1987
            Phys Lett {\bf A124} 215 
\bibitem {} Fernandez F M, Ma Q and Tipping R H 1989
            Phys Rev {\bf A39} 1605
\bibitem {} Znojil M 1997 J Math Phys {\bf 38} 5087 
\bibitem {} Znojil M 1996 Phys Lett {\bf A222} 291
\bibitem {} Lai C S 1981 Phys Rev {\bf A23} 455
\bibitem {} Bender C M and Orszag S A, " Advanced Mathematical Methods
for Scientists and Engineers" (McGraw - Hill, New York, 1978).
\bibitem {} Handy C R et al. 1993 J Phys {\bf A26} 2635
\bibitem {} Mustafa O and Barakat T 1997 Commun Theor Phys {\bf 28} 257 
\bibitem {} Mustafa O and Barakat T 1998 Commun Theor Phys {\bf 29} 587
\bibitem {} Barakat T, Odeh M and Mustafa O 1998 J Phys {\bf A31} 3469 
\bibitem {} Mustafa O 1993 J Phys; Condens. Matter {\bf 5} 1327 
\bibitem {} Mustafa O 1996 J Phys; Condens. Matter {\bf 8} 8073 
\bibitem {} Mustafa O and Chhajlany S C 1994 Phys Rev {\bf A50} 2926 
\end{thebibliography}
\end{document}